\documentstyle[11pt]{article}
\newcommand{\be}{\begin{equation}}
\newcommand{\ee}{\end{equation}}
\newcommand{\abz}{\hspace*{.5in}}
\newcommand{\nab}{\nabla}
\newcommand{\vphi}{{\bf \varphi}(x)}
\newcommand{\vphibar}{{\overline{\bf \varphi}}(x)}
\renewcommand{\thesection}{\Roman{section}}
\renewcommand{\theequation}{\arabic{section}.\arabic{equation}}

\refstepcounter{equation}

\begin{document}
\title{Conformal Transformations of the Wigner Function and Solutions of
the Quantum Corrected Vlasov Equation\thanks{gr--qc/9402015; submitted to
Clas. Quantum Grav.}}
\author{Oleg A. Fonarev\thanks{E--mail: OLG@vms.huji.ac.il}
\\ {\em Racah Institute of Physics, The Hebrew University}
\\ {\em Jerusalem 91904, Israel}}
\date{}

\maketitle
\begin{abstract}
We study conformal properties of the quantum kinetic equations
in curved spacetime.
A transformation law for the covariant Wigner function under conformal
transformations of a spacetime is derived by using the formalism of tangent
bundles.
The conformal invariance of the quantum corrected Vlasov equation is proven.
This provides a basis for generating new solutions of the quantum kinetic
equations in the presence of gravitational and other external fields.
We use our method to find explicit quantum corrections to the class of locally
isotropic distributions, to which equilibrium distributions belong.
We show that the quantum corrected stress--energy tensor for such distributions
has, in general, a non--equilibrium structure.
Local thermal equilibrium is possible in quantum systems only if an underlying
spacetime is conformally static (not stationary).
Possible applications of our results are discussed.
\end{abstract}

\thispagestyle{empty}
\setcounter{page}{0}
\newpage
\setcounter{page}{1}

\section{Introduction}
\setcounter{equation}{0}
\abz
In a regime where the curvature of a spacetime is less than the Planck scale,
a semiclassical approximation to the quantum gravity is considered to be
sufficient \cite{kn:birdav,kn:parker}.
This approximation is valid for "nearly" classical systems, for which  quantum
fluctuations in the stress-energy
tensor are small enough \cite{kn:ford}.
In such cases the method of Wigner functions \cite{kn:wigner,kn:carruthers,
kn:degroot} is particularly convenient.
It allows one to derive, from quantum field equations, quantum
kinetic equations for Wigner functions.
Iterative solutions of the kinetic equations give quantum corrections to
classical distribution functions, which are used to compute expectation
values of local observables such as the stress-energy tensor, the number--flux
vector etc.
The standard definition of the flat--space Wigner function exploits the
Wigner--
Weyl transformation \cite{kn:carruthers} which fails in curved spacetime.
Different approaches have been proposed to extend the notion of the
Wigner--Weyl
transformation to curved spacetime.
In Ref. \cite{kn:winter}, a Wigner--Weyl--type transformation is defined by
considering the bundle of geodesic lines on a manifold.
Ref. \cite{kn:hu} utilizes Riemann normal coordinate systems and adiabatic
expansions similar to one used by \cite{kn:bunch} for studying ultraviolet
properties of Green functions in curved spacetime.
In Ref. \cite{kn:fonSU}, a covariant Wigner function is defined by using the
formalism of tangent bundles.
All three approaches mentioned, being intrinsically different, are locally
equivalent to each other \cite{kn:fonJMP}, in the sense of
adiabatic expansions, and yield the same quantum corrected Vlasov--type
kinetic equations.
The mathematical structure of the equations is rather complicated, even at
lowest
adiabatic order.
This makes difficult the study of properties of generic solutions of the
quantum
kinetic equations in an arbitrary background spacetime.
No doubt that any explicit computation of the Wigner function may be useful for
better understanding of quantum kinetic processes in gravitating systems.\\
\abz
If a spacetime manifold is endued with some symmetries then both classical and
quantum kinetic equations can often be solved explicitly.
By now, only in a very few cases explicit solutions of the quantum corrected
Vlasov equation in curved spacetime have been obtained.
They include the Freidmann--Robertson--Walker cosmology \cite{kn:pirk,
kn:fonPL2}, stationary spacetimes \cite{kn:fonPL1} and some others
 \cite{kn:diss}.
A common property of those solutions is that the quantum corrections are local
in phase space.
Once a solution of the Vlasov equation (a classical distribution function)
is known, one can compute, to a given adiabatic order, the Wigner function
which corresponds to this solution, by applying a certain finite order
differential
operator to the classical distribution function.
This property implies that one can express the expectation value of the
stress--
energy tensor or of any other local observable in terms of moments of the
classical
distribution function multiplied by some local geometrical quantities.
This facilitates to a great extent the analysis of the back--reaction problem
in
those cases.\\ \abz
The natural question arises how symmetries of a spacetime
manifest themselves in Wigner functions.
In the present paper, we consider a particular, but important, type of
geometrical
symmetries, the conformal symmetry.
For simplicity, attention will be restricted to the case of a scalar field,
spin--$\frac{1}{2}$ and spin--1 fields can be treated similarly
\cite{kn:fonJMP,kn:diss}.
The conformal symmetry is usually understood as the invariance
of the generalized  massless Klein--Gordon equation
(with the term $\frac{1}{6} R \varphi$
involved \cite{kn:birdav}) under the
conformal rescaling of the metric and of the scalar field: $g_{\alpha\beta}
 \rightarrow a^{2} g_{\alpha\beta}, {\bf \varphi} \rightarrow {\bf \varphi}/a$.
If the scalar field is coupled to an external potential, the latter must also
be transformed in such a way that the scalar field equation remains invariant
under this extended group of conformal transformations. \\ \abz
Two different views at the conformal symmetry are possible.
When the external potential has a direct physical meaning, the conformal
symmetry is a {\it dynamical} symmetry of the physical system under
consideration.
This is the case, for example, in the $\varphi^{4}$ theory where the
expectation
 value $\langle {\bf \varphi}^{3} \rangle$ plays a role of an external
potential
for quantum perturbations.
The second point of view is formal.
One can treat, for example, the mass of the scalar field as an
external potential and use the conformal symmetry as a convenient tool for
generating new solutions of the field equation \cite{kn:tool}.
\\ \abz
In this paper, we study transformation properties of the covariant  Wigner
function under the conformal transformation.
As we have mentioned, different curved--space extensions
of the flat--space Wigner function are possible, all of them giving similar
adiabatic expansions.
The use of the tangent bundle technique is particularly convenient for
exploring mathematical properties of the Wigner function.
By using the formalism of tangent bundles we derive, to second adiabatic order,
the transformation law for the Wigner
function and prove the conformal
invariance of the quantum--corrected Vlasov equation.
We then use our result to
 find explicit quantum corrections to the class of distribution
functions which are locally isotropic in momentum space, i.e. have the form
${F_{cl}}(x,p) = {F}(x,{u^{\alpha}}(x) p_{\alpha})$.
As it was shown by \cite{kn:tauber,kn:ehlers}, in the case where external
fields other than gravitation are absent, such form of the distribution
function imposes the following restriction on the underlying spacetime: it must
be
or stationary or conformally static (with a special form of the conformal
factor).
In the presence of an external potential this condition has to be modified.
We do this by using the conformal properties of the Vlasov equation.
We also analyse the structure of the quantum corrected stress--energy tensor
and
number--flux vector, and show that the quantum corrections lead, in general, to
non--vanishing heat flux and viscosity, even if the classical distribution
implies local thermal equilibrium.\\
\abz
The paper is organized as follows.
In Sec.\ ~\ref{sec-definition}, we outline the formalism of the Wigner function
in curved spacetime.
Section ~\ref{sec-conformal} derives the transformation law for the Wigner
function.
The lowest order quantum corrections to the isotropic distributions are
evaluated
in Sec.\ ~\ref{sec-isotropic}, and the quantum corrected stress--energy tensor
and number--flux vector are calculated in Sec.\ ~\ref{sec-structure}.
Section ~\ref{sec-conclusion} contains some remarks concerning possible
applications
of our results.
In  ~\ref{sec-law}, technical details used for the evaluation of the
transformation law are presented, and in  ~\ref{sec-kinetic}, the conformal
invariance of the quantum corrected Vlasov equation is proven.
{}~\ref{sec-energy} recalls some properties of the stress--energy tensor.
\\ \abz
Conventions used throughout are the following.
Greek indices run from 0 to 3, Latin indices from 1 to 3. The signature for the
metric tensor is $(+---)$. The Riemann tensor $ R^{\mu}_{\nu\alpha\beta}$
is defined by
\be
\left[ \nab_{\alpha} , \nab_{\beta} \right] \,
X^{\mu} = R^{\mu}_{\nu\alpha\beta} \, X^{\nu} \; .
\ee
The Ricci tensor is $R_{\alpha\beta} = R^{\nu}_{\alpha\nu\beta}$,
the scalar curvature $R = R^{\alpha}_{\alpha}$.
\section{The covariant Wigner function}
\setcounter{equation}{0}
\abz \label{sec-definition}
To keep the paper self contained, we shall outline our definition and some
properties of the covariant Wigner function in curved spacetime \cite{kn:fonSU,
kn:fonJMP}.
Let $\vphi$ be a scalar field on a spacetime manifold  $\cal M$ with the metric
${g_{\alpha\beta}}(x)$, and  ${\cal T}_{x}(\cal M)$ be the tangent space at a
point
$x$.
The horizontal lift of the covariant derivative operator $\nab_{\alpha}$ to the
tangent
bundle ${\cal T} (\cal M)$ is defined by \cite{kn:yano}
\be
\hat{\nab}_{\alpha} = \nab_{\alpha} - \Gamma_{\alpha\nu}^{\beta} \: y^{\nu} \:
\frac{\partial}{\partial y^{\beta}} \; ,   \label{eq:dy}
\ee
where {$y^{\alpha}$} are components of a tangent vector pertaining to  ${\cal
T}_{x}
(\cal M)$ , and ${\Gamma_{\alpha\nu}^{\beta}}(x)$ are the Christoffel symbols.
Let us introduce the horizontal lift of the scalar field:
\be
{\bf \Phi}(x,y) = \exp(y^{\alpha} \hat{\nab}_{\alpha})\, \vphi \; .
\label{eq:Phi}
\ee
Then, the covariant Wigner function is defined as follows:
\be
{f}(x,p) = (\pi\hbar)^{-4}\: \sqrt{-{g}(x)}\:\int_{{\cal T}_{x}(\cal M)}
d^{4}\,y\:e^{-2iy^{\alpha}p_{\alpha}/\hbar}\:\langle {{\bf \Phi}}(x,-y){{\bf
\Phi}^{\dagger}}(x,y)\rangle \; , \label{eq:wigf}
\ee
where the bracket stand for an ensemble averaging with some density matrix
defined
on a Cauchy hypersurface \cite{kn:habib}, the dagger indicates the hermitian
conjugation.
\\ \abz
If the field $\vphi$ is a solution of the generalized Klein--Gordon equation:
\be
\left(\hbar^{2}\,\nab_{\alpha} \nab^{\alpha} - \frac{1}{6}\,\hbar^{2}\, {R}(x)
 + {V}(x) \right) \vphi = 0 \; , \label{eq:klein}
\ee
then the Wigner function obeys the quantum corrected Liouville--Vlasov equation
supplemented by the quantum corrected mass--shell constraint \cite{kn:winter,
kn:hu,kn:fonJMP}.
The explicit form of the equations to second adiabatic order is given in
 ~\ref{sec-kinetic}.
We note here that a Wigner function satisfying these equations (to second
adiabatic order) is represented as the series in derivatives of the
Dirac $\delta$--function \cite{kn:hu,kn:fonJMP}:
\be
{f}(x,p) = \left( {F_{cl}}(x,p) + \hbar^2 {F_{qu}}(x,p) \right) \,
{\delta}(\Omega) +
\hbar^2 {F_{1}}(x,p) \, {\delta'}(\Omega) +
\hbar^2 {F_{2}}(x,p) \, {\delta''}(\Omega) +
\hbar^2 {F_{3}}(x,p) \, {\delta'''}(\Omega) \; , \label{eq:deltas}
\ee
with the argument of the $\delta$--function being
\be
\Omega = {g^{\alpha\beta}}(x) p_{\alpha} p_{\beta} - {V}(x) \; .
\label{eq:omega}
\ee
The functions ${F_{n}}(x,p)$ , $n=1,2,3$, are expressed in terms of the
classical
distribution function $ {F_{cl}}(x,p)$, in such a way that the quantum
corrected
mass--shell constraint is satisfied (see ~\ref{sec-kinetic}), assuming
that  $ {F_{cl}}(x,p)$ obeys the Vlasov equation:
\be
{\delta}(\Omega) \left(\hat{{\cal L}}  {F_{cl}}(x,p) \right) = 0 \; .
\label{eq:vlasoveq}
\ee
Here $\hat{{\cal L}}$ is the Liouville--Vlasov operator \cite{kn:vlasov,
kn:israel,kn:stewart}:
\be
\hat{{\cal L}} = p^{\alpha} D_{\alpha} + \frac{1}{2} V_{,\alpha}
\frac{\partial}
{\partial p_{\alpha}}  \; , \label{eq:vlasovoperator}
\ee
$D_{\alpha}$ being the horizontal lift of the covariant derivative operator to
the
cotangent bundle \cite{kn:yano}:
\be
D_{\alpha} = \nab_{\alpha} + \Gamma_{\alpha\beta}^{\mu}\: p_{\mu}\:
\frac{\partial}{\partial p_{\beta}}  \; .  \label{eq:dp}
\ee
The function $ {F_{qu}}(x,p) $ is found by integrating the quantum transport
equation (see ~\ref{sec-kinetic})
along classical trajectories \cite{kn:hu,kn:fonJMP}.
\\ \abz
Equation (\ref{eq:vlasoveq}) defines the classical distribution function only
on the mass shell $\Omega = 0$.
This is sufficient so far as one is
interested in classical observables defined on the mass shell \cite{kn:israel}.
In quantum kinetic theory the situation is different \cite{kn:fonJMP}.
Though the function  $ {F_{cl}}(x,p)$ is tied up with the $\delta$--function
in the expansion (\ref{eq:deltas}), its properties out of the mass shell
influence upon both the off--shell quantum corrections, the $F_{n}$'s, and the
on--shell quantum corrections, $ {F_{qu}}(x,p) $.
The latter becomes evident if we get rid of the  $\delta$--function in Eq.\
(\ref{eq:vlasoveq}).
The function  $ {F_{cl}}(x,p)$ must satisfy the generalized Vlasov equation:
\be
\hat{{\cal L}} \: {F_{cl}}(x,p) = \Omega \, {\Delta_{F}}(x,p) \; ,
\label{eq:generalvlasov}
\ee
$\Delta_{F}$ being an arbitrary function which is assumed to be non--singular
on the mass shell.
Then the equation governing the evolution of  $ {F_{qu}}(x,p) $ explicitly
involves the function $\Delta_{F}$ (see ~\ref{sec-kinetic}).
In the next section we shall see that, in general, one can not get rid of
$\Delta_{F}$.
Even if the right--hand side of Eq.\ (\ref{eq:generalvlasov})
happens to vanish for some system, for systems conformally related to the one
under consideration, the
Liouville--Vlasov operator annihilates $F_{cl}$ on the mass shell only.
\section{The Wigner function and conformal transformations of a spacetime}
\setcounter{equation}{0}
\abz \label{sec-conformal}
Let us now consider the conformal transformation on a spacetime manifold
(we shall use throughout an overline for quantities in the conformally related
spacetime):
\be
\begin{array}{lll}
{g_{\alpha\beta}}(x) & = & {a}(x)^2 \, {\overline{g}_{\alpha\beta}}(x) \; ,\\
\vphi & = & \vphibar/{a}(x) \; ,\\
{V}(x) & = & {\overline{V}}(x)/{a}(x)^2 \; , \label{eq:conf}
\end{array}
\ee
${a}(x)$ being an arbitrary smooth function.
Equation (\ref{eq:klein}) is known to be invariant under such a transformation
 \cite{kn:birdav}.
If $\vphi$ is a solution of Eq. (\ref{eq:klein}) on the manifold ${\cal M}$
with
the metric ${g_{\alpha\beta}}(x)$ and in the presence of the external potential
${V}(x)$, then $ \vphibar$ is a solution of the conformally transformed
equation on the manifold $\overline{\cal M}$ with the metric
$ {\overline{g}_{\alpha\beta}}(x) $ and in the presence of the external
potential
$ {\overline{V}}(x)$, and vice versa.
\\ \abz
One can expect that the Wigner function ${f}(x,p)$ associated with the
field $\varphi$ must be somehow connected to the Wigner function
${\overline{f}}(x,p)$ associated with the field $\overline{\varphi}$.
In the classical limit ($\hbar \rightarrow 0$) the connection is very simple.
The mass shell (\ref{eq:omega}) is only scaled under the conformal
transformation
(we treat covariant components of the momentum vector, $p_{\alpha}$, and
contravariant components of the vector $y^{\alpha}$ as parameters which are not
changed under conformal transformations, therefore, making difference between
the tangent and cotangent spaces).
Thus,
\be
\Omega = \overline{\Omega}\, a^{-2} \; . \label{eq:deltabar}
\ee
Next, under the conformal transformation
(\ref{eq:conf}), the Liouville--Vlasov operator is transformed as follows:
\be
\hat{{\cal L}} =  \hat{\overline{\cal L}} \, a^{-2} +
 a_{\alpha} \, \frac{\partial}{\partial p_{\alpha}} \, \Omega \; ,
\label{eq:vlasovtrans}
\ee
where
\be
a_{\alpha} = \frac{\partial}{\partial x^{\alpha}} \ln a \; . \label{eq:da}
\ee
By comparing Eqs. (\ref{eq:deltabar}),(\ref{eq:vlasovtrans}) with Eqs.
(\ref{eq:masshell}) and (\ref{eq:transport}), one can conclude that
 the transformation law for the Wigner function should be
\be
{f}(x,p) = {a}(x)^2 \,{\overline{f}}(x,p) + \mbox{quantum corrections} \; .
\label{eq:transclass}
\ee
\abz
We could also arrive at Eq. (\ref{eq:transclass}) by writing the transformation
law for the Liouville--Vlasov operator in the form
\be
\hat{{\cal L}} = a^{-2}\, \hat{\overline{\cal L}} +
\Omega \, a_{\alpha} \, \frac{\partial}{\partial p_{\alpha}} \; .
\label{eq:vlasovtrans2}
\ee
{}From Eqs. (\ref{eq:vlasoveq}) and (\ref{eq:vlasovtrans2}) it follows that the
classical distribution  function remains invariant under the conformal
transformation.
Then Eqs. (\ref{eq:deltas}) and (\ref{eq:deltabar}) imply
(\ref{eq:transclass}).
\\ \abz
One can also see from Eq. (\ref{eq:vlasovtrans2}) that, unlike Eq.
(\ref{eq:vlasoveq}), the off--shell Vlasov
equation (\ref{eq:generalvlasov}) is not invariant under the conformal
transformation.
Namely, the function $\Delta_{F}$ is changed as follows:
\be
 {\Delta_{F}}(x,p) = {\overline{\Delta_{F}}}(x,p) + a_{\alpha}
\frac{\partial}{\partial p_{\alpha}}  {F_{cl}}(x,p) \; . \label{eq:delftrans}
\ee
\abz
In order to find the quantum corrections in Eq. (\ref{eq:transclass}), we must
turn to the definition of the covariant Wigner function (\ref{eq:wigf}).
First, we write the transformation law for the operator
$\hat{\nab}_{\alpha}$
, Eq. (\ref{eq:dy}), when acting to a scalar function in the tangent bundle.
Under the conformal transformation of the metric, Eq. (\ref{eq:conf}), it is
transformed as follows (cf. Eq. (\ref{eq:gammatrans})):
\be
\hat{\nab}_{\alpha} = \hat{\overline{\nab}}_{\alpha} - a_{\nu} y^{\nu}
\frac{\partial}{\partial y^{\alpha}} - a_{\alpha} y^{\nu} \frac
{\partial}{\partial y^{\nu}} + y_{\alpha} a^{\nu}
\frac{\partial}{\partial y^{\nu}} \; . \label{eq:nabtrans}
\ee
Let us now write the field (\ref{eq:Phi}) in the following way (compare Eq.
(\ref{eq:conf})):
\be
{\bf \Phi}(x,y) = \left( e^{(y^{\alpha} \hat{\nab}_{\alpha})} a^{-1}
e^{(- y^{\alpha} \hat{\nab}_{\alpha})} \right) \; \left(
e^{(y^{\alpha} \hat{\nab}_{\alpha})} \, e^{(- y^{\alpha}
\hat{\overline{\nab}}_{\alpha})} \right) \; \left(
e^{(y^{\alpha} \hat{\overline{\nab}}_{\alpha})} \vphibar \right) \; .
\label{eq:Phitrans}
\ee
The last factor on the right--hand--side of Eq. (\ref{eq:Phitrans}) is , by the
definition, ${\overline{\bf \Phi}}(x,y) $.
The second factor, which is an operator acting in the tangent bundle, can be
expanded as follows (see ~\ref{sec-law}):
\begin{eqnarray}
\hat{\overline{Z}}(x,y) & := & e^{(y^{\alpha} \hat{\nab}_{\alpha})} \,
e^{(- y^{\alpha} \hat{\overline{\nab}}_{\alpha})} \nonumber \\
& =& 1 + A^{\alpha} \, \left( \frac{1}{2} \hat{\nab}_{\alpha} -
\frac{\partial}{\partial y^{\alpha}} \right) + B^{\alpha}\, \left( \frac{1}{3}
\hat{\nab}_{\alpha} - \frac{1}{2}
\frac{\partial}{\partial y^{\alpha}}\right) \nonumber \\
\mbox{} & +& \frac{1}{2} A^{\alpha} A^{\beta}\,
\left( \frac{1}{4} \hat{\nab}_{\alpha}
\hat{\nab}_{\beta} - \frac{\partial}{\partial y^{\beta}} \hat{\nab}_{\alpha}
+ \frac{\partial^{2}}{\partial y^{\alpha}\partial y^{\beta}}\right) \nonumber
\\
\mbox{}& +& \mbox{terms of higher orders} \; , \label{eq:Z}
\end{eqnarray}
where $A^{\alpha}$ and $B^{\alpha}$ are given in ~\ref{sec-law}. \\ \abz
Lastly, the first factor on the right--hand side of Eq. (\ref{eq:Phitrans}) is
a
function which involves the conformal factor ${a}(x)$ and its derivatives.
The known formula gives the expansion:
\begin{eqnarray}
{A}(x,y)& :=&   e^{(y^{\alpha} \hat{\nab}_{\alpha})} a^{-1}
e^{(- y^{\alpha} \hat{\nab}_{\alpha})}  \nonumber \\
\mbox{}& =& a^{-1} \left( 1 - y^{\alpha} a_{\alpha} + \frac{1}{2} (y^{\alpha}
a_{\alpha})^{2} - \frac{1}{2} y^{\alpha} y^{\beta} a_{\alpha ; \beta}
\right) \nonumber \\
\mbox{} &+&  \mbox{terms of higher orders} \; , \label{eq:Axy}
\end{eqnarray}
where $ a_{\alpha ; \beta}:= \nab_{\beta} a_{\alpha}$.
On the right--hand side of Eqs. (\ref{eq:Z}), (\ref{eq:Axy}), we  have kept
 only those terms which yield lowest order quantum corrections to Eq.
(\ref{eq:transclass}).
\\ \abz
We are now prepared to write down the transformation law for the Wigner
function, correct to second adiabatic order.
Before doing that we note the following identity \cite{kn:fonJMP}:
\be
\frac{\partial}{\partial y^{\alpha}} \, {{\bf \Phi}}(x,y) =
  \hat{\nabla}_{\alpha} \, {{\bf \Phi}}(x,y) \\
\mbox{} + \mbox{terms of second or higher orders} \; . \label{eq:dphi}
\ee
Then, with the aid of the results of ~\ref{sec-law}, we finally get:
\begin{eqnarray}
{f}(x,p)& =& a^{2} \left( 1 + \frac{\hbar^{2}}{4}\, a_{\alpha ; \beta} \,
\frac{\partial^{2}}{\partial p_{\alpha} \partial p_{\beta}}
+ \frac{\hbar^{2}}{8}\, {A^{\alpha}}(\partial_{p})\,
\overline{D}_{\alpha}
 - \frac{\hbar^{2}}{24}\, {B^{\alpha}}(\partial_{p}) \,
p_{\alpha} \right)\, {\overline{f}}(x,p) \nonumber \\
 \mbox{} &+& \mbox{terms of higher adiabatic order} \; , \label{eq:ftrans}
\end{eqnarray}
where the operator $\overline{D}_{\alpha}$ is conformally related to
(\ref{eq:dp}), and
${A^{\alpha}}(\partial_{p})$ and
$ {B^{\alpha}}(\partial_{p})$ are the following operators
(cf. Eqs. (\ref{eq:A}),(\ref{eq:B})):
\be
{A^{\alpha}}(\partial_{p}) = ( 2 \, a_{\mu}\, \delta_{\nu}^{\alpha} -
\overline{a}^{\alpha}\, \overline{g}_{\mu\nu} )\,
\frac{\partial^{2}}{\partial p_{\mu} \partial p_{\nu}} \; , \label{eq:Ap}
\ee
\be
{B^{\alpha}}(\partial_{p}) = ( 2 \overline{a}_{\mu ; \nu}\,
\delta_{\gamma}^{\alpha} -  \overline{a}^{\alpha}_{;\mu}\,
 \overline{g}_{\nu\gamma} - 8 a_{\mu}\, a_{\nu}\, \delta_{\gamma}^{\alpha} +
2\, \overline{a}^{\beta}\, a_{\beta} \, \overline{g}_{\mu\nu} \,
\delta_{\gamma}^{\alpha} + 4 \, \overline{a}^{\alpha} \, a_{\mu} \,
\overline{g}_{\nu\gamma} ) \,
\frac{\partial^{3}}{\partial p_{\mu} \partial p_{\nu} \partial p_{\gamma}}
\; . \label{eq:Bp}
\ee
We have denoted $ \overline{a}_{\mu ; \nu} := \overline{\nab}_{\nu} a_{\mu}$
and
$\overline{a}^{\alpha} := \overline{g}^{\alpha\beta} a_{\beta}$.
\\ \abz
Equations (\ref{eq:ftrans})--(\ref{eq:Bp}) give the transformation law for the
Wigner function which we have been seeking, correct to second adiabatic order.
It is worth noting that the transformation law has been derived irrespective
of any  specific form of quantum kinetic equations, and it holds independently
of whether or not the Wigner functions satisfy these equations.
One could follow a different way.
Given equations which govern the evolution of a Wigner function, one may seek
a transformation which leaves the equations invariant under the conformal
transformation of the metric and the external potential.
This is what we did in order to obtain the transformation law at zeroth order,
Eq. (\ref{eq:transclass}).
However, this program meets serious problems at higher orders because the
structure of the quantum kinetic equations in curved spacetime is very
complicated.
Instead in ~\ref{sec-kinetic} we prove, to second order, the conformal
invariance
of the collisionless quantum kinetic equations by using the transformation law
for the Wigner function. \\
\abz
One may make sure that Eqs. (\ref{eq:ftrans})--(\ref{eq:Bp}) yield the right
transformation law for physical observables.
For example, the number--flux vector defined by \cite{kn:fonJMP}
\be
\langle {\bf J}_{\alpha}(x) \rangle = \int\frac{d^{4}\,p}{\sqrt{-{g}(x)}}\:
p_{\alpha}\: {f}(x,p) \; , \label{eq:current}
\ee
is transformed as follows:
\be
(-g)^{1/4}\, \langle {\bf J}_{\alpha} \rangle =
(- \overline{g})^{1/4} \, \langle \overline{\bf J}_{\alpha} \rangle  \; .
\label{eq:curtrans}
\ee
The transformation law (\ref{eq:curtrans}) implies that the number of
"particles" (defined as in kinetic theory) in a comoving three--volume, $dN :=
\langle {\bf J}^{\alpha} \rangle d \Sigma_{\alpha}$, is an invariant of the
group of conformal transformations.
Here $ d \Sigma_{\alpha} := \sqrt{-g}\, \varepsilon_{\alpha\sigma\mu\nu} \,
dx^{\sigma}dx^{\mu} dx^{\nu}$. \\
\abz
The transformation law for the stress--energy tensor depends upon an explicit
form of the external potential.
For a scalar field coupled to curvature one gets (see ~\ref{sec-energy})
\be
(g/\overline{g})^{1/4} \, \langle {\bf T}_{\alpha\beta} \rangle =
 \langle \overline{{\bf T}}_{\alpha\beta} \rangle + \frac{\hbar^{2}}{2} \,
(6 \xi -1)\, \left( 2 \, a \, a_{(\alpha} \, \overline{\nab}_{\beta)} - a\,
\overline{g}_{\alpha\beta} \, a_{\nu} \, \overline{\nab}^{\nu} -
\overline{g}_{\alpha\beta} \, \overline{a}^{\nu}_{\nu}
\right)\, (\overline{N}/a) \; , \label{eq:energytrans}
\ee
where $\xi$ is the nonminimal gravitational coupling constant and
\be
{\overline{N}}(x) = \int\frac{d^{4}\,p}{\sqrt{-{\overline{g}}(x)}}\:
{\overline{f}}(x,p) \; . \label{eq:Nbar}
\ee
\section{Quantum corrections to isotropic distributions}
\setcounter{equation}{0}
\abz \label{sec-isotropic}
Let us consider a system in external fields $g_{\alpha\beta}$ and $V$ whose
classical distribution function (which is assumed obeying the Vlasov equation)
is locally isotropic in momentum space, that is ${F_{cl}}(x,p) = {F}(x,
{u^{\alpha}}(x) p_{\alpha})$, ${u^{\alpha}}(x)$ being some world velocity
field.
Of course, the external fields must be consistent with the symmetry of the
distribution function.
{}From the results of Refs. \cite{kn:tauber,kn:ehlers} and from the
conformal invariance of the Vlasov equation it follows that only two
possibilities exist:
\\
A. The metric is conformally stationary, i.\ e.\ the line element takes the
form
(in preferred coordinates)
\be
d s^{2} = a^{2} ( d t^{2} + 2 \overline{g}_{0 i} d x^{0} d x^{i} +
\overline{g}_{ij} d x^{i} d x^{j} ) \; , \label{eq:ds}
\ee
with $ \overline{g}_{0 i}$ and $\overline{g}_{ij}$ being functions of the
spatial
coordinates {$x^{i}$} alone, and either the potential $V$ equals zero or the
function $\overline{V} := V a^{2}$ does not depend on time; \\
B. The metric is conformally static, that is $ \overline{g}_{0 i} = 0$ in
some coordinate system, and the function  $\overline{V}$ depends on time
alone
(since the Vlasov equation is linear in $V$, one can conclude that in Case B
 the function $\overline{V}$ may be of the form \cite{kn:ehlers}
$\overline{V} = {\overline{V}_{1}}(t) +  {\overline{V}_{2}}(x^{i})$ ;
we shall not deal with this more general situation here).
\\ \abz
In accordance with our strategy, we can solve the quantum kinetic equation in
an auxiliary spacetime with the line element
\be
d \overline{s}^{2} = d t^{2} + 2 \overline{g}_{0 i} d x^{0} d x^{i} +
\overline{g}_{ij} d x^{i} d x^{j} \; , \label{eq:dsbar}
\ee
and then use Eq. (\ref{eq:ftrans}) to calculate the Wigner function in the
physical spacetime.
\\
\abz
In the spacetime with the metric $\overline{g}_{\alpha\beta}$ there exist a
unit time--like Killing vector $\overline{\xi}^{\alpha}$,
$\overline{\xi}^{\alpha} \overline{\xi}_{\alpha} = 1$ ($\overline{\xi}^{\alpha}
=
\delta_{0}^{\alpha}$ in the preferred coordinates), which obeys the equation
\cite{kn:yano2}:
\begin{eqnarray}
\overline{\nab}_{\alpha} \overline{\xi}_{\beta}  +
\overline{\nab}_{\beta} \overline{\xi}_{\alpha} = 0 & \mbox{in Case A} \; ,
\label{eq:killingstation} \\
  \mbox{or} \hspace{.1in}
\overline{\nab}_{\alpha} \overline{\xi}_{\beta} = 0 & \mbox{in Case B} \; .
\label{eq:killingstatic}
\end{eqnarray}
The potential must satisfy the conditions:
\begin{eqnarray}
\overline{\xi}^{\alpha} \overline{V}_{; \alpha} = 0 =
\overline{\xi}^{\alpha} \overline{\xi}^{\beta} \overline{V}_{; \alpha\beta}
= \ldots & \mbox{in Case A} \; ,
\label{eq:vstation} \\
\overline{\Delta}^{\alpha\beta}  \overline{V}_{; \alpha} = 0 =
\overline{\Delta}^{\alpha\beta}  \overline{V}_{; \alpha\beta}
 = \ldots & \mbox{in Case B} \; .
\label{eq:vstatic}
\end{eqnarray}
Here
\be
\overline{\Delta}^{\alpha\beta} := \overline{\xi}^{\alpha}
\overline{\xi}^{\beta} - \overline{g}^{\alpha\beta} \label{eq:projection}
\ee
is the space--like projection operator and $ \overline{V}_{; \alpha} :=
 \overline{\nab}_{\alpha} \overline{V}$ etc. \\
\abz
In the two cases considered, the locally isotropic distribution
functions \cite{kn:ehlers} take the form
(one can see that in Case B the distribution function bellow has the desired
form $ {F}(x,{u^{\alpha}}(x) p_{\alpha})$ on the mass shell):
\begin{eqnarray}
F_{cl} = {F}(\overline{\xi}^{\alpha} p_{\alpha})  & \mbox{in Case A} \; ,
\label{eq:fstation} \\
F_{cl} = {F}(\sqrt{\overline{\Delta}^{\alpha\beta}  p_{\alpha} p_{\beta}})
 & \mbox{in Case B} \; . \label{eq:fstatic}
\end{eqnarray}
\abz
We proceed now to the derivation of the quantum corrections to the classical
distribution functions.
Both the function (\ref{eq:fstation}) and the function (\ref{eq:fstatic})
satisfy the off--shell Vlasov equation
in the external fields $\overline{g}_{\alpha\beta}$ and
$\overline{V}$, with $\overline{\Delta}_{F}$ being zero (see Eq.
(\ref{eq:generalvlasov})).
Note, however, that for systems conformally related to the ones under
consideration, $\Delta_{F} \neq 0$,
as it follows from Eq. (\ref{eq:delftrans}). \\
\abz
The semiclassical Wigner functions which correspond to the distribution
functions (\ref{eq:fstation}) and (\ref{eq:fstatic}) in the spacetime
$\overline{\cal M}$ are represented by the same expansions as in Eq.
(\ref{eq:deltas}), with the argument of the $\delta$--function being
\be
\overline{\Omega} = \overline{g}^{\alpha\beta} p_{\alpha} p_{\beta} -
\overline{V} \; . \label{eq:omegabar}
\ee
The off--shell quantum corrections are obtained by substituting the classical
distribution functions (\ref{eq:fstation}) and (\ref{eq:fstatic}) into Eq.
(\ref{eq:Fs}).
Both in Case A and in Case B, they can be written in the form:
\begin{eqnarray}
\overline{F}_{1} & = & - \frac{1}{6}\, \left( \overline{R} + 2
\overline{R}^{\mu \; \nu}_{\alpha . \beta} \, p_{\mu} p_{\nu}
\, \frac{\partial^{2}}{\partial p_{\alpha}
\partial p_{\beta}} \right)\, F_{cl} \; , \nonumber \\
\overline{F}_{2}& = & \left( - \frac{1}{3} \, \overline{R}^{\mu\nu} \,
p_{\mu} p_{\nu} +
\frac{1}{4}\, \overline{V}^{; \alpha}_{\alpha} + \frac{1}{2}\,
\overline{V}^{; \mu}_{\nu} \, p_{\mu}  \frac{\partial}{\partial p_{\nu}}
\right) \, F_{cl} \; ,\nonumber \\
\overline{F}_{3} &= & \frac{1}{12} \, \left( 2\, \overline{V}^{; \mu\nu}\,
p_{\mu} p_{\nu}
 -  \overline{V}^{; \alpha} \overline{V}_{; \alpha} \right) \, F_{cl} \; .
\label{eq:Fstat}
\end{eqnarray}
\abz
Next, the on--shell quantum corrections are found from Eq. (\ref{eq:Fqu}).
It follows from the conditions listed in Eqs. (\ref{eq:vstation}) and
(\ref{eq:vstatic}) that
the equation for $\overline{F}_{qu}$ can be integrated as if the potential
$\overline{V}$ were constant.
A solution of this equation which corresponds to the classical distribution
(\ref{eq:fstation}) was found in Ref. \cite{kn:fonPL1}.
It reads (with $\overline{\xi}^{\alpha} \overline{\xi}_{\alpha} = 1$)
\begin{eqnarray}
\overline{F}_{qu} = - \frac{1}{6} \, \overline{R}_{\alpha\beta} \,
\frac{\partial^{2}}{\partial p_{\alpha} \partial p_{\beta}} \, F_{cl} &
\mbox{in Case A.} \label{eq:FquA}
\end{eqnarray}
\abz
Let us now consider Case B.
It follows from Eq. (\ref{eq:killingstatic}) that the following identities hold
 \cite{kn:yano2}:
\be
\overline{\xi}^{\alpha} \, \overline{R}_{\alpha\beta\mu\nu} = 0 =
\overline{\xi}^{\sigma} \, \overline{R}_{\alpha\beta\mu\nu ; \sigma} \; .
\label{eq:Rstatic}
\ee
By using Eqs. (\ref{eq:killingstatic}), (\ref{eq:vstatic}) and
(\ref{eq:Rstatic})
one readily gets that the function
\begin{eqnarray}
\overline{F}_{qu} = - \frac{1}{12} \, \overline{R}_{\alpha\beta} \,
\frac{\partial^{2}}{\partial p_{\alpha} \partial p_{\beta}}\,  F_{cl} &
\mbox{in Case B} \label{eq:FquB}
\end{eqnarray}
satisfies the quantum kinetic equation (\ref{eq:Fqu}), with $F_{cl}$ being of
the form (\ref{eq:fstatic}).
\\
\abz
Equations (\ref{eq:Fstat}) and (\ref{eq:FquA}) give the second order quantum
corrections to the
locally isotropic distribution (\ref{eq:fstation}) in a stationary spacetime in
the presence of a static potential (Case A), while Eqs. (\ref{eq:Fstat}) and
(\ref{eq:FquB}) give the quantum corrections to the distribution
(\ref{eq:fstatic}) in a static spacetime in the presence of a homogeneous
potential (Case B).
As we have said, the Wigner functions for systems conformally related to the
ones which have been considered in this section are
obtained by using Eqs. (\ref{eq:conf}) and (\ref{eq:ftrans}).
In the next section we shall analyse the structure of the quantum corrected
number--flux vector and stress--energy tensor in these cases.
\section{Quantum corrections to a perfect fluid}
\setcounter{equation}{0}
\abz \label{sec-structure}
The distributions (\ref{eq:fstation}) and (\ref{eq:fstatic}) both imply that
the classical stress--energy tensor has a perfect fluid structure
 \cite{kn:ehlers},
\be
\langle \overline{{\bf T}}_{cl}^{\alpha\beta} \rangle =
( \overline{\varepsilon}_{cl} + \overline{P}_{cl} ) \, \overline{\xi}^{\alpha}
\, \overline{\xi}^{\beta} -  \overline{P}_{cl}\, \overline{g}^{\alpha\beta} \;
,
\label{eq:stressenergyperfect}
\ee
the classical energy density $\overline{\varepsilon}_{cl}$ and pressure
$\overline{P}_{cl}$ being given by
\begin{eqnarray}
\overline{\varepsilon}_{cl} = 2 \pi \int_{0}^{\infty} d\, p \: p^{2} \:
(\overline{V} +
p^{2})^{1/2}\: {F}_{cl}(\overline{V},p) \; , \nonumber \\
\overline{P}_{cl} = \frac{1}{3}\, 2 \pi \int_{0}^{\infty} d\, p \: p^{4} \:
(\overline{V} +
p^{2})^{-1/2} \: {F}_{cl}(\overline{V},p) \; ,
\label{eq:energyperfect}
\end{eqnarray}
with (compare Eqs. (\ref{eq:fstation}), (\ref{eq:fstatic}))
\be
 {F}_{cl}(\overline{V},p) = \left\{ \begin{array}{ll}
{F}(\sqrt{\overline{V} +p^{2}}) & \mbox{in Case A} \\
{F}(p) & \mbox{in Case B} \; .\end{array} \right.  \label{eq:fperfect}
\ee
The classical number--flux vector associated with the locally isotropic
distributions reads:
\be
\langle \overline{\bf J}_{cl}^{\alpha} \rangle = \overline{n}_{cl} \,
\overline{\xi}^{\alpha} \; , \label{eq:currentperfect}
\ee
the classical number density being
\be
\overline{n}_{cl} = 2 \pi \int_{0}^{\infty} d\, p \: p^{2} \:{F}_{cl}
(\overline{V},p) \; .
\label{eq:numberperfect}
\ee
\abz
It should be mentioned that we have taken into account positive "energies"
only,
$\overline{E} = \sqrt{\overline{V} +p^{2}}$, when integrating over the mass
shell.
This is tantamount to multiplying the distribution function by the step
function \cite{kn:stewart}
${\Theta}(\overline{\xi}^{\alpha} p_{\alpha})$.
Though $\overline{\xi}^{\alpha} p_{\alpha}$ is not a constant of motion in Case
B,
the step function is "ignored" by the Liouville--Vlasov operator, neither it
contributes to the quantum kinetic equation, since differentiation of the step
function gives ${\delta}(\overline{\xi}^{\alpha} p_{\alpha})$ which vanishes on
the mass shell (we assume that, in the classical limit, $\overline{V}$ is
non--negative everywhere in the spacetime).
Note that negative "energies", $\overline{E} = - \sqrt{\overline{V} +p^{2}}$,
correspond to antiparticles \cite{kn:degroot}.
\\ \abz
The quantum corrections to Eqs.
(\ref{eq:stressenergyperfect}),(\ref{eq:currentperfect})
are easily obtained by substituting the Wigner functions found in the preceding
section into Eqs. (\ref{eq:stressenergy}) and (\ref{eq:current}).
We shall first consider Case B.
The second order quantum corrections to the number flux
(\ref{eq:currentperfect}) can be shown to vanish.
The quantum corrected stress--energy tensor can be written, after some simple
algebra, in the following form:
\begin{eqnarray}
\langle \overline{{\bf T}}^{\alpha\beta} \rangle =
\langle \overline{{\bf T}}_{cl}^{\alpha\beta} \rangle +
\hbar^{2}\, (\xi - \frac{1}{6}) \, \overline{R}^{\alpha\beta} \, M_{1,1} +
\hbar^{2} \, \ddot{\overline{V}} \,
\overline{\Delta}^{\alpha\beta}\, \left(\frac{1}{2}\,
(\xi - \frac{1}{6})\, M_{1,2} - \frac{1}{24}\,\overline{V}\, M_{1,3}\right)
\nonumber \\
\mbox{} -
\hbar^{2} \, \dot{\overline{V}}^{2}\, \overline{\Delta}^{\alpha\beta}\, \left(
 \frac{3}{4} \,
(\xi - \frac{1}{6})\, M_{1,3} - \frac{1}{96}\,  M_{2,4} - \frac{1}{16} \,
\overline{V}\,M_{1,4} \right) +
\frac{1}{32} \, \dot{\overline{V}}^{2}\, \overline{\xi}^{\alpha}\,
\overline{\xi}^{\beta} \, M_{1,3} \; . \label{eq:stressenergyB}
\end{eqnarray}
Here $\dot{\overline{V}} := \overline{\xi}^{\alpha} \, \overline{\nab}_{\alpha}
\overline{V}$ etc.\ , $\overline{\Delta}^{\alpha\beta}$ is defined by Eq.
(\ref{eq:projection}), and the M's denote the integrals:
\be
M_{n,k} =  2 \pi \int_{0}^{\infty} d\, p \: p^{2n} \: (\overline{V} +
p^{2})^{1/2- k}\: {F}(p) \; . \label{eq:Mnk}
\ee
\abz
The stress--energy tensor (\ref{eq:stressenergyB}) has an "almost"  perfect
fluid
form, except the second term which involves $\overline{R}^{\alpha\beta} $.
The latter has in general off--diagonal components.
Note, however, that the Einstein equations impose severe restrictions on the
underlying spacetime.
Suppose, for example, that the conformal factor in Eq. (\ref{eq:ds}) depends on
time only: $a={a}(t)$ (recall that we are considering Case B, that is
$ \overline{g}_{0 i} = 0$ in the preferred coordinates).
If the total stress--energy tensor of classical matter is assumed to have a
perfect fluid form then the physical spacetime is necessary a Robertson--Walker
one \cite{kn:ehlers}.
Then $\overline{R}^{\alpha\beta} = 2 K\, \overline{\Delta}^{\alpha\beta}$, with
$K = -1, 0, 1$, that is the quantum--corrected stress--energy tensor preserves
the perfect fluid structure in this case. \\ \abz
Let us write down the stress--energy tensor with the second order quantum
corrections in a Robertson--Walker spacetime with the line element
\be
d s^{2} = a^{2}\, ( d t^{2} - \frac{d r^2}{1 - K r^2} - r^2 d \theta^{2} -
r^2 \sin^{2}\theta d \varphi^{2} ) \; . \label{eq:RW}
\ee
Since the Bianchi identities hold automatically for the stress--energy tensor
derived in kinetic theory, it suffices to know the energy density $\varepsilon
:= \langle T_{0}^{0} \rangle$.
Making use of Eqs. (\ref{eq:energytrans}), (\ref{eq:energyperfect}) and
(\ref{eq:stressenergyB}) gives
\begin{eqnarray}
\varepsilon & = & \frac{1}{a^{4}}\, M_{1,0} + \frac{\hbar^{2}}{2 a^{4}} \,
( 1 - 6 \xi) \,
(\frac{\dot{a}^{2}}{a^{2}} - K)\, M_{1,1} \nonumber \\
\mbox{} + \frac{\hbar^{2}}{4}\, ( 1 & -& 6 \xi)\, \frac{\dot{a}}{a^{3}}\,
( \dot{U} +
2 \, \frac{\dot{a}}{a}\, U)\, M_{1,2} + \frac{\hbar^{2}}{32}\, ( \dot{U} +
2\, \frac{\dot{a}}{a}\, U)^{2}\, M_{1,3} \; . \label{eq:energyRW}
\end{eqnarray}
Here dots stand for the time derivative, and the M's are given by Eq.
(\ref{eq:Mnk}), with $\overline{V} = U a^{2}$, $U$ being the classical part of
the
external potential in the Robertson--Walker spacetime (see ~\ref{sec-energy}).
Equation (\ref{eq:energyRW}) extends the result of Ref. \cite{kn:pirk} to the
case of a variable potential and the nonminimal coupling to curvature.
Physical manifestations of the quantum kinetic corrections in the
Friedmann--Robertson--Walker cosmology have been examined in Ref.
\cite{kn:fonPL2}.
It was shown there that the quantum corrected energy density causes an
apparent time variation of the value of the gravitational constant in the early
universe.
\\ \abz
We turn now to Case A of the preceding section.
The structure of the quantum corrections is more complicated in this case.
To avoid lengthy expressions, we shall confine our attention to systems for
which $V = 0$.
Substituting the distribution function (\ref{eq:fstation}) and the quantum
corrections to it given by Eqs. (\ref{eq:Fstat}), (\ref{eq:FquA})
into Eqs. (\ref{eq:current}) and
(\ref{eq:stressenergy}) gives the quantum corrected number--flux vector and
stress--energy tensor in a stationary spacetime (recall that
$\overline{\xi}^{\alpha} \overline{\xi}_{\alpha} = 1$):
\be
\langle \overline{\bf J}^{\alpha} \rangle = \overline{\xi}^{\alpha}\, M_{2} -
\frac{\hbar^{2}}{6}\,  \overline{\Delta}^{\alpha\mu}\, \overline{\xi}^{\nu}
\,\overline{R}_{\mu\nu}\, M_{0} \; , \label{eq:currentA}
\ee
\begin{eqnarray}
\langle \overline{{\bf T}}^{\alpha\beta} \rangle& = & \frac{1}{3}\, (4\,
\overline{\xi}^{\alpha}\, \overline{\xi}^{\beta} - \overline{g}^{\alpha\beta})
\, (M_{3} - \frac{\hbar^{2}}{2} \, \overline{\xi}^{\mu} \, \overline{\xi}^{\nu}
\, \overline{R}_{\mu\nu} \, M_{1}) \nonumber \\
\mbox{} & +&  \frac{2 \hbar^{2}}{3} \, ( \overline{\xi}^{(\alpha} \,
\overline{R}^{\beta) \mu} \, \overline{\xi}_{\mu} - \overline{\xi}_{\mu} \,
\overline{\xi}_{\nu} \,\overline{R}^{\alpha\mu\beta\nu}) \, M_{1} \; .
\label{eq:energyA}
\end{eqnarray}
Here $\overline{R}^{\alpha\mu\beta\nu}$ is the Riemann tensor of the stationary
spacetime, and
\be
M_{n} =  2 \pi \int_{0}^{\infty} d\, p \: p^{n} \: {F}(p) \; . \label{eq:Mn}
\ee
\abz
The remarkable feature of the quantum corrections in Eqs. (\ref{eq:currentA}),
(\ref{eq:energyA}) is that they have a non--equilibrium structure.
To show this explicitly, let us compute the invariants
of the number--flux vector and stress--energy tensor.
The hydrodynamical velocity associated with the number flux (\ref{eq:currentA})
is
\be
\overline{v}^{\alpha} = \overline{\xi}^{\alpha} - \frac{\hbar^{2}}{6}\,
\overline{\Delta}^{\alpha\mu} \, \overline{\xi}^{\nu} \,
\overline{R}_{\mu\nu} \, M_{0}/M_{2} \; . \label{eq:v}
\ee
If we decompose the stress--energy tensor (\ref{eq:energyA}) like in Eq.
(\ref{eq:energygeneral}) we find next, with the aid of Eq.
(\ref{eq:energydeviation}), that the eigenvector of
$\langle \overline{{\bf T}}^{\alpha\beta} \rangle$ is
\be
\overline{u}^{\alpha} = \overline{\xi}^{\alpha} - \frac{\hbar^{2}}{4} \,
\overline{\Delta}^{\alpha\mu}\, \overline{\xi}^{\nu} \,
\overline{R}_{\mu\nu}\, M_{1}/M_{3} \; . \label{eq:u}
\ee
In general, $\overline{u}^{\alpha}$ does not coincide with
$\overline{v}^{\alpha}$.
This implies that the heat flux defined by \cite{kn:israel,kn:degroot}
\be
\overline{q}^{\alpha} = (\overline{\varepsilon} + \overline{P}) \,
(\overline{u}^{\alpha} - \overline{v}^{\alpha}) \label{eq:heat}
\ee
does not vanish.
Indeed, substituting Eqs. (\ref{eq:v}),(\ref{eq:u}) into (\ref{eq:heat}) gives
\be
\overline{q}^{\alpha} = \frac{\hbar^{2}}{9} \, ( 2 M_{0} M_{3}/M_{2} - M_{1})\,
\overline{\Delta}^{\alpha\mu} \,\overline{\xi}^{\nu} \,
\overline{R}_{\mu\nu} \; . \label{eq:heatA}
\ee
The combination of the moments on the right--hand side of Eq. (\ref{eq:heatA})
can be zero for a very special choice of the function ${F}(p)$.
It is surprising that the heat flux as well as the viscosity
(see ~\ref{sec-energy}) do not vanish even when the classical distribution
function describes local thermal equilibrium \cite{kn:fonwig}.
Though the effect is purely quantum, it could play an important role in the
early universe and in the evolution of massive stars.
Some physical consequences of this will be considered elsewhere.
\\ \abz
If (and only if) the manifold $\overline{\cal M}$ is static, that is the
classical fluid motion is irrotational, the quantum corrections in Eqs.
(\ref{eq:currentA}),(\ref{eq:energyA}) vanish, which is consistent with Case B
in the limit $\overline{V} \rightarrow 0$, $\xi = 1/6$  (compare Eq.
(\ref{eq:stressenergyB}).
\\ \abz
Finally, the quantum corrected number--flux vector and stress--energy tensor
in a conformally stationary spacetime can be easily found with the aid of Eqs.
(\ref{eq:curtrans}),(\ref{eq:energytrans}).
For the conformal coupling ($\xi = 1/6$) they are obtained by simply
multiplying Eqs. (\ref{eq:currentA}) and (\ref{eq:energyA}), respectively,
by $a^{-4}$ and $a^{-6}$.
\section{Concluding remarks}
\abz \label{sec-conclusion}
In summary, we have considered conformal properties of the covariant Wigner
function.
The transformation law (\ref{eq:ftrans}) relates Wigner functions in manifolds
conformally related to each other, correct to second adiabatic order.
Given a solution of quantum kinetic equations which govern the evolution of a
quantum distribution function (a Wigner function) of a specific system, one
is able to evaluate Wigner functions for a wide class of systems conformally
related to the one under consideration.
\\ \abz
There exist two possible applications of the result.
First, our method allows one to simplify an original system by reducing it to
a simpler one for which the evolution can be easily solved.
As an example, we have found explicit solutions of the collisionless quantum
kinetic equations in conformally stationary/static spacetimes and analyzed
the structure of the quantum corrected stress--energy tensor and number--flux
vector. \\ \abz
Second, in some theories of gravity, like Brans--Dicke--Jordan--type ones which
have been attracting a great deal of attention  in recent years in connection
with the
low energy limit of string theory (see, for instance, Ref. \cite{kn:tseytlin}),
 two
metrics are involved: a physical metric which enters matter equations, and the
so called Einstein metric which satisfies Einstein--like equations.
The two metrics are usually conformally related to each other (though more
general relations have been suggested, see Ref. \cite{kn:bekenstein}), and one
might want to solve matter equations on the physical manifold and then express
the solutions in terms of the metric associated with gravity.
Our method described in the present paper is especially suitable for such
theories. \\ \abz
We would like to finish our discussion by the following remark.
It is well known \cite{kn:tauber,kn:ehlers,kn:israel,kn:stewart} that local
thermal equilibrium in a relativistic gas is
possible in a very restrictive class of physical spacetimes.
As we have shown, in a quantum system thermal equilibrium is even more
exceptional than in a classical one.
Spacetimes in which quantum fields can be in local thermal equilibrium are
restricted to conformally static ones.
Although we have analyzed collisionless equations only, this is
also correct for collision--dominated systems \cite{kn:stewart}, because
quantum corrections to the collision integral are supposed to be small compare
with ones to the Liouville--Vlasov operator when a system is nearly
equilibrium.
Thus our results provide further motivation for studying non--equilibrium
processes in quantum systems in curved spacetime.

\section*{Acknowledgements}
\abz
I am grateful to Professors L. P. Horwitz and J. D. Bekenstein for helpful
discussions of the work.
\appendix
\renewcommand{\thesection}{Appendix \Alph{section}}
\renewcommand{\theequation}{\Alph{section}\arabic{equation}}
\section{: Evaluation of the transformation law for the Wigner function}
\setcounter{equation}{0}
\abz \label{sec-law}
One can derive Eq. (\ref{eq:ftrans}) by applying the
Campbell--Baker--Hausdorff formula \cite{kn:serre}:
\be
\exp(\hat{D}) \: \exp(- \hat{\overline{D}}) = \exp\left(\hat{D} -
\hat{\overline{D}}
- \frac{1}{2} [\hat{D},\hat{\overline{D}}] + \frac{1}{12} [\hat{D} +
\hat{\overline{D}},[\hat{D},\hat{\overline{D}}]] + \ldots \right) \; ,
\label{eq:hausdorff}
\ee
for the operators $y^{\alpha} \nab_{\alpha}$ and $y^{\alpha} \overline{\nab}_
{\alpha}$.
Making use of Eq. (\ref{eq:nabtrans}),(\ref{eq:hausdorff}) and taking into
account the fact \cite{kn:fonJMP} that the operator $ \nab_{\alpha}$
commutes with $\frac{\partial}{\partial y^{\alpha}}$ and annihilates
$y^{\alpha}$ immediately give Eq. (\ref{eq:Z}) with
\be
A^{\alpha} = 2 y^{\alpha} y^{\nu} a_{\nu} - y^{\nu} y_{\nu} a^{\alpha} \; ,
\label{eq:A}
\ee
\be
\begin{array}{c}
B^{\alpha} =  y^{\mu} \hat{\overline{\nab}}_{\mu} A^{\alpha} - A^{\mu}
\frac{\partial}{\partial y^{\mu}} A^{\alpha} \nonumber \\
\mbox{} = 2 y^{\alpha} y^{\mu} y^{\nu} (\overline{\nab}_{\mu} a_{\nu}) -
y^{\mu} y^{\nu} y_{\nu} (\overline{\nab}_{\mu} a^{\alpha}) - 8 y^{\alpha}
(y^{\nu} a_{\nu})^{2} + 2 y^{\alpha} y^{\nu} y_{\nu} a^{\mu} a_{\mu} +
4 y^{\nu} y_{\nu} y^{\mu} a_{\mu} a^{\alpha} \; . \label{eq:B}
\end{array}
\ee
Next, with the aid of Eqs. (\ref{eq:Z}) and (\ref{eq:dphi}) we obtain:
\begin{eqnarray}
\hat{\overline{Z}}(x,y) {\overline{\bf \Phi}}(x,y) & = & \left( 1 - \frac{1}{2}
A^{\alpha} \hat{\overline{\nab}}_{\alpha} + \frac{1}{6}  B^{\alpha}
\frac{\partial}{\partial y^{\alpha}} + \frac{1}{8} A^{\alpha} A^{\mu}
\hat{\overline{\nab}}_{\alpha}  \hat{\overline{\nab}}_{\mu} \right) \:
{\overline{\bf \Phi}}(x,y) \nonumber \\
\mbox{} &+& \mbox{terms of third or higher orders} \; . \label{eq:Zphi}
\end{eqnarray}
Using Eqs. (\ref{eq:Phitrans}),(\ref{eq:Axy}),(\ref{eq:Zphi}) and
analogous identities for the conjugate field, we arrive at
\begin{eqnarray}
{\bf \Phi}(x,-y){{\bf \Phi}^{\dagger}}(x,y) = a^{-2} \left( 1 -  y^{\alpha}
y^{\beta} a_{\alpha ; \beta} - \frac{1}{2} A^{\alpha} \hat{\overline{\nab}}_
{\alpha} - \frac{1}{6}  B^{\alpha}
\frac{\partial}{\partial y^{\alpha}} \right. \nonumber \\
\left. \mbox{} + \frac{1}{8} A^{\alpha} A^{\mu}
\hat{\overline{\nab}}_{\alpha}  \hat{\overline{\nab}}_{\mu} \right) \:
\left({\overline{\bf \Phi}}(x,-y) {\overline{\bf \Phi}^{\dagger}}(x,y)
\right) +
\mbox{terms of higher orders} \; . \label{eq:phiphi}
\end{eqnarray}
Finally, substituting Eq. (\ref{eq:phiphi}) into the definition of the Wigner
function
(\ref{eq:wigf}) and integrating by parts give Eq. (\ref{eq:ftrans})
(the last term in the bracket on the right--hand side of Eq. (\ref{eq:phiphi})
leads to terms of fourth adiabatic order which we do not keep).

\section{: Conformal invariance of the collisionless quantum kinetic equations}
\setcounter{equation}{0}
\abz \label{sec-kinetic}
The equations for a Wigner function of a scalar field obeying the generalized
Klein--Gordon equation (\ref{eq:klein}) read \cite{kn:winter,kn:hu,kn:fonJMP},
to second adiabatic order:
\begin{eqnarray}
\hat{{\cal L}} \: {f}(x,p) = \hbar^{2} \hat{\Lambda}
{f}(x,p) \; , \label{eq:transport}\\
\Omega \: {f}(x,p) = - \hbar^{2} \hat{\Pi}
{f}(x,p) \; , \label{eq:masshell}
\end{eqnarray}
where $\Omega$ and $\hat{{\cal L}}$ are respectively given by Eqs.
(\ref{eq:omega})
 and (\ref{eq:vlasovoperator}), and the operators on the right--hand side of
Eqs.
 (\ref{eq:transport}) and {\ref{eq:masshell} are
\begin{eqnarray}
\hat{\Lambda}& =&  \frac{1}{6} R_{\nu\beta\mu\alpha} p^{\mu}
\frac{\partial^{2}}{\partial p_{\alpha} \partial p_{\beta}}
D^{\nu} - \frac{1}{24}
R_{\alpha\mu\beta\nu ; \sigma} p^{\mu} p^{\nu} \frac{\partial^{3}}{\partial
p_{\alpha} \partial p_{\beta} \partial p_{\sigma}} \nonumber \\
\mbox{} + \frac{1}{12} R_{\alpha}^{\nu} \frac{\partial}{\partial p_{\alpha}}
D_{\nu}
 &-& \frac{1}{24} R_{\alpha\beta ; \nu} p^{\nu} \frac{\partial^{2}}{\partial
p_{\alpha} \partial p_{\beta}} - \frac{1}{24} R_{; \alpha}
\frac{\partial}{\partial p_{\alpha}} + \frac{1}{48} V_{; \alpha\beta\sigma}
\frac{\partial^{3}}{\partial
p_{\alpha} \partial p_{\beta} \partial p_{\sigma}} \; , \label{eq:lambda}
\end{eqnarray}
\begin{eqnarray}
\hat{\Pi} = &-& \frac{1}{4}\, D^{\alpha} D_{\alpha} - \frac{1}{6} R -
\frac{1}{12} R_{\alpha\mu\beta\nu}
p^{\mu} p^{\nu} \frac{\partial^{2}}{\partial p_{\alpha} \partial p_{\beta}}
\nonumber \\
\mbox{} &-&
\frac{1}{4} R_{\mu\nu} p^{\mu} \frac{\partial}{\partial p_{\nu}} +
\frac{1}{8} V_{; \alpha\beta} \frac{\partial^{2}}{\partial p_{\alpha}
\partial p_{\beta}} \; . \label{eq:pi}
\end{eqnarray}
Here $D_{\alpha}$ is defined by Eq. (\ref{eq:dp}), semicolons signify the
covariant differentiation associated with the metric $g_{\mu\nu}$.
\\
\abz
The operators $\hat{\Lambda}$ and $\hat{\Pi}$ satisfy the following
identity \cite{kn:fonJMP}:
\be
[\hat{\Lambda}, \Omega] = [\hat{{\cal L}}, \hat{\Pi}] \; , \label{eq:lambdapi}
\ee
which allows one to look for solutions to Eqs. (\ref{eq:transport}),
(\ref{eq:masshell}) of the form (\ref{eq:deltas}) \cite{kn:hu,kn:fonJMP}.
The F's in Eq. (\ref{eq:deltas}) are given by
\begin{eqnarray}
{F_{1}}(x,p) & = & \hat{\Pi} \: {F_{cl}}(x,p) \; , \nonumber \\
{F_{2}}(x,p) & = & \frac{1}{2} \, [\hat{\Pi},\Omega] \: {F_{cl}}(x,p) \; ,
\nonumber \\
{F_{3}}(x,p) & = & \frac{1}{6} \, [[\hat{\Pi},\Omega],\Omega] \: {F_{cl}}(x,p)
\; ,
\label{eq:Fs}
\end{eqnarray}
while the function ${F_{qu}}(x,p)$ satisfies the equation \cite{kn:fonJMP}:
\be
\hat{{\cal L}}\:{F_{qu}}(x,p) = \hat{\Lambda}\: {F_{cl}}(x,p) +
\hat{\Pi} \: {\Delta_{F}}(x,p) \; , \label{eq:Fqu}
\ee
${\Delta_{F}}(x,p) $ being defined by Eq. (\ref{eq:generalvlasov}). \\ \abz
Consider now a conformally related spacetime with the metric
$\overline{g}_{\alpha
\beta} = g_{\alpha\beta} / a^{2}$ and suppose that ${\overline{f}}(x,p)$
satisfies the quantum corrected Vlasov equation and mass--shell constraint in
the
 presence of the external fields $\overline{g}_{\alpha\beta}$ and $\overline{V}
= V a^{2}$.
As we have shown in Sec. ~\ref{sec-conformal},
the conformal transformation of the spacetime induces the following
transformation of the Wigner function defined in the cotangent bundle:
\be
{f}(x,p) = a^2 \, (1 + \hbar^{2} \, \hat{C})\, {\overline{f}}(x,p) \; ,
\label{eq:ftrans1}
\ee
where the operator $\hat{C}$ is, to second adiabatic order (cf. Eq.
(\ref{eq:ftrans}),
\begin{eqnarray}
\hat{C} = \frac{1}{24} \left(- 4 \overline{a}_{\mu;\nu} +
\overline{a}^{\alpha}_
{;\alpha} \overline{g}_{\mu\nu} + 28 a_{\mu} a_{\nu} - 10 \overline{a}^{\alpha}
a_{\alpha} \overline{g}_{\mu\nu}
+ 6 a_{\mu} \overline{D}_{\nu} - 3
\overline{g}_{\mu\nu} \overline{a}^{\alpha} \overline{D}_{\alpha} \right) \:
\frac{\partial^{2}}{\partial p_{\mu} \partial p_{\nu}} \nonumber \\
\mbox{} + \frac{1}{24} \left(
- 2 \overline{a}_{\mu;\nu} p_{\sigma} + \overline{a}^{\alpha}_{; \sigma}
\overline{g}_{\mu\nu} p_{\alpha}
+ 8 a_{\mu} a_{\nu}p_{\sigma} - 2 \overline{a}^{\alpha}
a_{\alpha} \overline{g}_{\mu\nu} p_{\sigma} - 4 \overline{g}_{\mu\nu}
a_{\sigma}
\overline{a}^{\alpha} p_{\alpha} \right) \: \frac{\partial^{3}}{\partial
p_{\mu} \partial p_{\nu} \partial p_{\sigma}} \; . \label{eq:C}
\end{eqnarray}
\abz
If one takes into account the transformation laws for the Wigner function, Eq.
 (\ref{eq:ftrans1}), and for the Liouville--Vlasov operator, Eq.
(\ref{eq:vlasovtrans}), and uses the equations for the function
${\overline{f}}(x,p)$,
one arrives at the following equations:
\begin{eqnarray}
\hat{{\cal L}} {f}(x,p) &=& \hbar^{2} \left(\hat{\overline{\Lambda}} +
[\hat{\overline{\cal L}}, \hat{C}] - a_{\alpha}
\frac{\partial}{\partial p_{\alpha}}\:
(\hat{\overline{\Pi}} + [\hat{C},\overline{\Omega}]) \right) \:
{\overline{f}}(x,p) \nonumber \\
\mbox{} &+& \mbox{terms of higher orders} \; , \label{eq:transporttrans}
\end{eqnarray}
\begin{eqnarray}
\Omega \: {f}(x,p) &=& - \hbar^{2} \left(\hat{\overline{\Pi}} + [\hat{C},
\overline{\Omega}] \right) \: {\overline{f}}(x,p)  \nonumber \\
\mbox{} &+& \mbox{terms of higher orders} \; . \label{eq:masstrans}
\end{eqnarray}
Now, our task is to prove that these equations are tantamount, to second
adiabatic
order, to Eqs. (\ref{eq:transport}), (\ref{eq:masshell}).
Tedious but straightforward calculations give the following results:
\be
\hat{\overline{\Pi}} + [\hat{C},\overline{\Omega}] = \hat{\Pi} \: a^{2} +
a_{\alpha} \frac{\partial}{\partial p_{\alpha}}\: \hat{\overline{\cal L}}
+ \frac{1}{4} (4 a_{\mu} a_{\nu} - \overline{a}^{\alpha}
a_{\alpha} \overline{g}_{\mu\nu} - \overline{a}_{\mu;\nu} ) \:
\frac{\partial^{2}}{\partial p_{\mu} \partial p_{\nu}}
\: \overline{\Omega} \; , \label{eq:Pitrans}
\ee
\begin{eqnarray}
\hat{\overline{\Lambda}} +
[\hat{\overline{\cal L}}, \hat{C}] &=& ( \hat{\Lambda} +
a_{\alpha} \frac{\partial}{\partial p_{\alpha}}\: \hat{\Pi} )\: a^{2}
+ \frac{1}{4} (\overline{a}_{\mu;\nu} + \overline{a}^{\alpha}
a_{\alpha} \overline{g}_{\mu\nu} ) \:
\frac{\partial^{2}}{\partial p_{\mu} \partial p_{\nu}} \:
\hat{\overline{\cal L}} \nonumber \\
\mbox{} &-& \frac{1}{24} (\overline{a}_{\mu;\nu\sigma} -
8 \overline{a}_{\mu;\nu}
a_{\sigma} + 4 \overline{a}^{\alpha} \overline{a}_{\mu;\alpha} - 6
\overline{a}^{\alpha} a_{\alpha} a_{\mu} \overline{g}_{\nu\sigma} )
\frac{\partial^{3}}{\partial p_{\mu} \partial p_{\nu}\partial p_{\sigma}}
\: \overline{\Omega}  \; . \label{eq:Lambdatrans}
\end{eqnarray}
When deriving Eqs. (\ref{eq:Pitrans}), (\ref{eq:Lambdatrans}) we have used the
transformation laws for the Christoffel symbols and Riemann tensor
{\cite{kn:yano2}:
\be
\Gamma_{\mu\nu}^{\alpha} - \overline{\Gamma}_{\mu\nu}^{\alpha} =
\delta_{\mu}^{\alpha} a_{\nu} + \delta_{\nu}^{\alpha} a_{\mu} -
\overline{g}_{\mu\nu} \overline{a}^{\alpha} \; , \label{eq:gammatrans}
\ee
\be
a^{-2} R_{\alpha\mu\beta\nu} - \overline{R}_{\alpha\mu\beta\nu} =
2 a_{\nu[\alpha} \overline{g}_{\mu]\beta} - 2 a_{\beta[\alpha} \overline{g}_
{\mu]\nu} \; , \label{eq:Rtrans}
\ee
where $a_{\alpha\beta} := \overline{a}_{\alpha;\beta} - a_{\alpha} a_{\beta}
+ \frac{1}{2} \,
\overline{g}_{\alpha\beta} \, \overline{a}^{\nu} a_{\nu}$.
Recall also that $a_{\alpha} := \partial_{\alpha} \ln a$, $\overline{a}^
{\alpha} := \overline{g}^{\alpha\beta} a_{\beta}$ and semicolons in Eqs.
(\ref{eq:C}), (\ref{eq:Pitrans})--(\ref{eq:Rtrans}) signify covariant
differentiation associated with the metric $\overline{g}_{\alpha\beta}$.
 \\ \abz
Substituting Eqs. (\ref{eq:Pitrans}), (\ref{eq:Lambdatrans}) into Eqs.
(\ref{eq:transporttrans}), (\ref{eq:masstrans}) completes the proof.
\section{: The stress--energy tensor}
\setcounter{equation}{0}
\abz \label{sec-energy}
Consider, for simplicity, the theory of a real scalar field described by the
action:
\be
S = \frac{1}{2} \int d^{4}\, x \:\sqrt{-g}\:(\hbar^{2} \nab^{\alpha}
{\bf \varphi} \nab_{\alpha}
{\bf \varphi} + \hbar^{2} \xi R {\bf \varphi}^{2} - U {\bf \varphi}^{2}) \; ,
\label{eq:action}
\ee
$R$ being the Ricci scalar, $U$ an external potential not involving a metric
dependence. \\ \abz
The variation of the action with respect to the metric tensor yields the
stress--
energy tensor \cite{kn:birdav}:
\begin{eqnarray}
{\bf T_{\mu\nu}}(\xi,V) = \hbar^{2} \nab_{\mu} {\bf \varphi} \nab_{\nu} {\bf
\varphi}
&+& \hbar^{2} \xi R_{\mu\nu} {\bf \varphi}^{2} \nonumber \\
\mbox{} - \frac{1}{2} g_{\mu\nu} ( \hbar^{2} \nab^{\alpha}
{\bf \varphi} \nab_{\alpha}
{\bf \varphi} + \frac{1}{6} \hbar^{2} R {\bf \varphi}^{2} - V {\bf
\varphi}^{2})
&-& \hbar^{2} \xi (\nab_{\mu} \nab_{\nu} - g_{\mu\nu} \nab^{\alpha}
\nab_{\alpha}) {\bf \varphi}^{2} \; . \label{eq:quantumstress}
\end{eqnarray}
We have introduced, for convenience, the generalized external potential
{\cite{kn:parker2}
\be
V = U - \hbar^{2} (\xi - \frac{1}{6}) R \; . \label{eq:VU}
\ee
The expectation value of the stress--energy tensor can be expressed in terms of
the covariant Wigner function (\ref{eq:wigf}) as follows \cite{kn:fonJMP}:
\begin{eqnarray}
\langle{\bf T_{\mu\nu}}(\xi,V)\rangle =  \int\frac{d^{4}\,p}{\sqrt{-{g}(x)}}\:
(p_{\mu} p_{\nu} + \hbar^{2} \xi R_{\mu\nu}) \:{f}(x,p) \nonumber \\
\mbox{} + \hbar^{2} (\frac{1}{4} - \xi) (\nab_{\mu} \nab_{\nu} - g_{\mu\nu}
\nab^{\alpha} \nab_{\alpha}) \int\frac{d^{4}\,p}{\sqrt{-{g}(x)}}\: {f}(x,p) \;
{}.
\label{eq:stressenergy}
\end{eqnarray}
\abz
The action (\ref{eq:action}) is invariant under the conformal transformation
(\ref{eq:conf}).
Notice that the potential $U$ is transformed as the following:
\be
U = \overline{U}/a^{2} + \hbar^{2}\, (1 - 6 \xi) \, a^{-3} \,
\overline{\nab}^{\alpha}
\overline{\nab}_{\alpha} a \; . \label{eq:utrans}
\ee
Since the transformation law for the potential explicitly involves the
dependence
on the metric, the stress--energy tensor is not conformally invariant.
With the aid of Eq. (\ref{eq:ftrans}) we obtain the following relation:
\begin{eqnarray}
 a^{2} \langle {\bf T}_{\mu\nu} \rangle & - & \langle
\overline{{\bf T}}_{\mu\nu}
\rangle  = \nonumber \\
\hbar^{2}\, (6 \xi -1) \, \left( a_{(\mu} \overline{\nab}_{\nu)} -
\frac{1}{2}\, \overline{g}_{\mu\nu} \, a_{\alpha} \overline{\nab}^{\alpha} -
 a_{\mu}  a_{\nu} \right. & -&\left. \frac{1}{2} \, \overline{g}_{\mu\nu}\, (
\overline{\nabla}^{\alpha}
 a_{\alpha}) \right)  \int\frac{d^{4}\,p}{\sqrt{-{\overline{g}}(x)}}\:
{\overline{f}}(x,p) \; . \label{eq:energytrans1}
\end{eqnarray}
Simple algebra then leads to Eq. (\ref{eq:energytrans}).
When deriving Eq. (\ref{eq:energytrans1}) we have used the identity
(\ref{eq:Rtrans}) and the following property of the operator (\ref{eq:dp}):
\be
\int\frac{d^{4}\,p}{\sqrt{-g}}\: D_{\alpha}\: {f}(x,p) = \nab_{\alpha} \:
\int\frac{d^{4}\,p}{\sqrt{-g}}\:{f}(x,p) \; , \label{eq:Dnab}
\ee
which was proven in Ref. \cite{kn:fonJMP}.
\\ \abz
{}From Eq. (\ref{eq:energytrans1}) it follows that the stress--energy tensor
(multiplied by the density $(-g)^{1/4}$) is
conformally invariant, but not tracefree, if $\xi = 1/6$.
$T_{\alpha}^{\alpha}$ vanishes only if $V = 0$.
\\ \abz
Any physical stress--energy tensor can be decomposed as follows
\cite{kn:degroot}:
\be
T_{\mu\nu} = (\varepsilon + P) u_{\mu} u_{\nu} - P g_{\mu\nu} + \Pi_{\mu\nu} \;
,
\label{eq:energygeneral}
\ee
where $\varepsilon$ and $P$ are, respectively, the energy density and the
isotropic pressure, $\Pi_{\mu\nu}$ is the tracefree viscosity tensor [$\Pi^
{\alpha}_{\alpha} = 0 = u^{\alpha} \Pi_{\alpha\beta}$], and
$u_{\nu}$ is a unit time--like
 vector [$u^{\alpha} u_{\alpha} = 1$].
\\ \abz
The perfect fluid stress--energy tensor takes the form:
\be
\stackrel{0}{T}_{\mu\nu} = (\stackrel{0}{\varepsilon} + \stackrel{0}{P})
\stackrel{0}{u}_{\mu} \stackrel{0}{u}_{\nu} -
\stackrel{0}{P}  g_{\mu\nu} \; . \label{eq:oT}
\ee
For small deviations from the perfect fluid structure, $T_{\mu\nu} =
\stackrel{0}{T}_{\mu\nu} + \stackrel{1}{T}_{\mu\nu}$, one easily finds,
to first order in the deviations:
\be
\begin{array}{lll}
\varepsilon & =&  \stackrel{0}{\varepsilon} +  u^{\alpha} u^{\beta}
\stackrel{1}{T}_{\alpha
\beta} \\
P & =&   \stackrel{0}{P} + \frac{1}{3} \Delta^{\alpha\beta}
\stackrel{1}{T}_{\alpha
\beta} \\
u_{\mu} & = & \stackrel{0}{u}_{\mu} - \Delta_{\mu}^{\alpha}  u^{\beta}
\stackrel{1}{T}_{\alpha
\beta}/(\stackrel{0}{\varepsilon} + \stackrel{0}{P}) \\
 \Pi_{\mu\nu} & = & (\Delta_{\mu}^{\alpha} \Delta_{\nu}^{\beta} - \frac{1}{3}
\Delta_{\mu\nu} \Delta^{\alpha\beta})\stackrel{1}{T}_{\alpha
\beta} \; , \label{eq:energydeviation}
\end{array}
\ee
where the projection operator has been introduced:
\be
\Delta_{\mu\nu} =  \stackrel{0}{u}_{\mu} \stackrel{0}{u}_{\nu} -  g_{\mu\nu} \;
{}.
\label{eq:uproject}
\ee

\end{document}